\documentstyle[preprint,tighten,aps,amstex,graphicx,times,floats]{revtex}

\begin{document}

\thispagestyle{empty}

\marginparwidth 1.cm
\setlength{\hoffset}{-1cm}
\newcommand{\mpar}[1]{{\marginpar{\hbadness10000%
                      \sloppy\hfuzz10pt\boldmath\bf\footnotesize#1}}%
                      \typeout{marginpar: #1}\ignorespaces}
\def\mda{\mpar{\hfil$\downarrow$\hfil}\ignorespaces}
\def\mua{\mpar{\hfil$\uparrow$\hfil}\ignorespaces}
\def\mla{\marginpar[\boldmath\hfil$\rightarrow$\hfil]%
                   {\boldmath\hfil$\leftarrow $\hfil}%
                    \typeout{marginpar: $\leftrightarrow$}\ignorespaces}

\def\ba{\begin{eqnarray}}
\def\ea{\end{eqnarray}}
\def\bq{\begin{equation}}
\def\eq{\end{equation}}

\renewcommand{\abstractname}{Abstract}
\renewcommand{\figurename}{Figure}
\renewcommand{\refname}{Bibliography}

\newcommand{\eg}{{\it e.g.}\;}
\newcommand{\ie}{{\it i.e.}\;}
\newcommand{\etal}{{\it et al.}\;}
\newcommand{\ibid}{{\it ibid.}\;}

\newcommand{\mx}{M_{\rm SUSY}}
\newcommand{\pt}{p_{\rm T}}
\newcommand{\et}{E_{\rm T}}
\newcommand{\del}{\varepsilon}
\newcommand{\sla}[1]{/\!\!\!#1}
\newcommand{\fb}{\;{\rm fb}}
\newcommand{\pb}{\;{\rm pb}}
\newcommand{\mev}{\;{\rm MeV}}
\newcommand{\gev}{\;{\rm GeV}}
\newcommand{\tev}{\;{\rm TeV}}
\newcommand{\abi}{\;{\rm ab}^{-1}}
\newcommand{\fbi}{\;{\rm fb}^{-1}}

\newcommand{\lsusy}{\lambda'_{\rm SUSY}}
\newcommand{\llq}{\lambda'_{\rm LQ}}
\newcommand{\msbar}{\overline{\rm MS}}
\newcommand{\met}{E\hspace{-0.45em}|\hspace{0.1em}}

\newcommand{\SP}{\scriptscriptstyle}
\newcommand{\stl}{\tilde{t}_{\SP L}}
\newcommand{\str}{\tilde{t}_{\SP R}}
\newcommand{\ste}{\tilde{t}_1}
\newcommand{\stz}{\tilde{t}_2}
\newcommand{\st}{\tilde{t}}
\newcommand{\gt}{\tilde{g}}
\newcommand{\sle}{\tilde{\tau}_1}
\newcommand{\slz}{\tilde{\tau}_2}
\newcommand{\che}{\tilde{\chi}^\pm_1}
\newcommand{\cpe}{\tilde{\chi}^+_1}
\newcommand{\cme}{\tilde{\chi}^-_1}
\newcommand{\nne}{\tilde{\chi}^0_1}
\newcommand{\nnz}{\tilde{\chi}^0_2}
\newcommand{\mse}{m_{\tilde{t}_{\SP 1}}}
\newcommand{\msz}{m_{\tilde{t}_{\SP 2}}}
\newcommand{\mst}{m_{\tilde{t}}}
\newcommand{\mg}{m_{\tilde{g}}}
\newcommand{\ms}{m_{\tilde{q}}}
\newcommand{\mt}{m_t}
\newcommand{\mheavy}{m_{\rm heavy}}
\newcommand{\mle}{m_{\tilde{\tau}_{\SP 1}}}
\newcommand{\mce}{m_{\tilde{\chi}^+_{\SP 1}}}
\newcommand{\mne}{m_{\tilde{\chi}^0_{\SP 1}}}

\preprint{
\font\fortssbx=cmssbx10 scaled \magstep2
\hbox to \hsize{
\hfill\vtop{\hbox{\bf MADPH-02-1308} 
            \hbox{\bf CERN-TH/2002-298}
            \hbox{\today}                    } }
}

\title{ 
Supersymmetric Dark Matter - How Light Can the LSP Be?  
}

\author{
Dan Hooper$^1$ and Tilman Plehn$^{1,2}$
} 

\address{ 
$^1$ Physics Department, University of Wisconsin, Madison, US \\
$^2$ Theory Division, CERN, Geneva, Switzerland} 

\maketitle 

\begin{abstract}
 Using a very minimal set of theoretical assumptions we derive a lower
 limit on the LSP mass in the MSSM. We only require that the LSP be
 the lightest neutralino, that it be responsible for the observed
 relic density and that the MSSM spectrum respect the LEP2 limits. We
 explicitly do not require any further knowledge about the MSSM
 spectrum or the mechanism of supersymmetry breaking. Under these
 assumptions we determine a firm lower limit on the neutralino LSP
 mass of $18\gev$. We estimate the effect of improved limits on the
 cold dark matter relic density as well as the effects of improved
 LEP2-type limits from a first stage of TESLA on the allowed range of
 neutralino LSP masses.
\end{abstract} 

\vspace{0.2in}


\section{Introduction}
\label{sec:intro}

One of the most puzzling experimental observations in astrophysics has
for a long time been the dominance of the invisible cold dark matter
in the universe. Experiments measuring the cosmic microwave
background~\cite{cmb}, high red-shift supernovae~\cite{sn}, galactic
clusters and galactic rotation curves~\cite{rotation} have found that
the matter density of the universe is $\Omega_M \simeq
0.3$--$0.4$. In contrast, constraints from big-bang nucleosynthesis
indicate that the baryonic matter density must be well below this
number~\cite{bbn}. Last but but not least, the observed density of
luminous matter is very small, $\Omega_L < 0.01$~\cite{luminous}.
Therefore, the vast majority of the mass in the universe is dark. In
addition, cosmic microwave background studies and large scale structure
formation requires that the majority of the dark matter be cold
(non-relativistic)~\cite{cmb,structure}.\bigskip

Starting from a completely different set of experiments and looking at
collider-oriented high energy physics, the challenge for the next
generation of experiments is to understand how electroweak symmetry is
broken and masses are generated. Electroweak precision
studies~\cite{lep_ewwg} clearly point into the direction of
spontaneous electroweak symmetry breaking --- a mechanism which
creates masses for fermions and gauge bosons but which also
automatically yields a scalar Higgs boson. Furthermore, precision
experiments prefer a light Higgs boson with a mass below $\sim
250\gev$. Unfortunately, the mass of a light Higgs boson in the
Standard Model is not stable in perturbation theory, but this mass
hierarchy problem can be naturally solved by adding supersymmetry to
the gauge symmetries on which the Standard Model is based. We
emphasize that supersymmetry does not accidentally cancel the
divergences in the Higgs boson mass nor does it have a complicated
array of mechanisms to get rid of them. The hierarchy problem is
solved by the most basic idea of a supersymmetry between fermionic and
bosonic fields~\cite{hls}.\smallskip

The most general supersymmetric Lagrangean, however, induces
flavor-changing neutral interactions which are experimentally very
well constrained~\cite{fcnc}. The simplest way to avoid these
constraints is an exact or approximate $R$ symmetry which translates
into the conservation of a supersymmetric spectrum quantum
number~\cite{rparity}. Although being inspired by flavor physics
constraints, this $R$ symmetry has a huge impact on astrophysics: it
leads to the existence of a stable lightest supersymmetric particle
(LSP). One possible experimental signature of LSPs with masses of the
order of the weak scale could be the measured amount of cold dark
matter. Which MSSM particle the LSP is depends on the
model parameters. Again there might be other theories with a discrete
symmetry which for that very reason lead to cold dark
matter~\cite{extrad}. However, in the MSSM the existence of the LSP is
not at all ad-hoc but the natural consequence of flavor physics
constraints.

\section{Supersymmetric Dark Matter}
\label{sec:dm}

If supersymmetry is to provide us with a suitable dark matter
candidate, we can say some things about the nature of the LSP. First,
it must be colorless and neutral to avoid
observation~\cite{non_cdm}. Although it may be possible for a colored
or charged particle to form bound states with Standard Model
particles, searches for exotic isotopes have ruled out exotic charged
bound states over a large mass range~\cite{isotopes}. Neutral exotic
bound states, consisting of squarks or gluons and Standard Model
particles, could possibly evade this type of detection, but would also
need to be very heavy or be carefully designed to evade collider
searches.  On the other hand these collider searches usually assume
that these strongly interacting superpartners decay to a weakly
interacting LSP~\cite{tev_limits,run2report}.
\smallskip

A second class of constraints on a SUSY dark matter candidate comes
from the limits placed by direct elastic scattering experiments.
Sneutrinos with relatively large elastic scattering cross sections can
be probed by these experiments. By now, all of sneutrino LSP parameter
space has been ruled out by direct and indirect
searches~\cite{sneutrinos}.\smallskip

This leaves the lightest neutralino as the neutral and colorless SUSY
dark matter candidate~\cite{neutralino}. Generally, it has a small
enough elastic scattering cross section to be missed by experimental
searches.  Moreover, the mass and annihilation cross section for a
neutralino LSP lie naturally in a region which yields a thermal dark
matter relic density in agreement with observation. In these LSP
annihilation processes light superpartners, such as scalar tau leptons,
play an important role as $t$ channel propagators.  We calculate the
neutralino LSP relic density using the full cross section, including
all resonances and thresholds, and solving the Boltzmann equation
numerically~\cite{darksusy,boltzman}.  The issue of neutralino
co-annihilation with other light superpartners will be discussed in
detail later in this letter.

\section{Theoretical Assumptions}
\label{sec:ass}

Of all supersymmetric parameters, the mass difference between the LSP
and heavier MSSM states has a particularly crucial impact on collider searches at
hadron colliders as well as at TESLA. The question we attempt to
answer in this paper is straightforward: how light can the LSP be
assuming nothing about the unknown SUSY breaking mechanism? \medskip

A huge number of constraints on the MSSM spectrum have been
accumulated over the last years. Using these constraints, observables
like the light Higgs boson mass, the LEP limits on chargino and
slepton masses, the squark and gluino mass limits from the Tevatron
and flavor physics constraints like the $b \to s\gamma$ rate can be
translated into limits on the LSP mass. However, all of these links
rely on theoretical assumptions, usually on the assumption of unified
Majorana fermion or scalar masses at some GUT scale, as it is
suggested by the gauge coupling unification~\cite{msugra,cdm_tesla}.
A very instructive discussion of these issues can be found, for
example, in Ref.~\cite{favorite}. More recently, some effort has been
put into the effect which breaking the weak gaugino mass and the
gluino mass unification can have on the detection of supersymmetric
dark matter~\cite{non_gluino}.  Speculative relations between MSSM
parameters can for example stem from the attempt to link features of
an unknown underlying string theory to the current experimental
results~\cite{whacko}. Top-down approaches to the MSSM spectrum can
only be a first step to understand the interplay between different
assumptions and observables. For example the effect of non-universal
Higgs masses should be and has been explored in detail~\cite{nuhm}. A
preliminary scan over the supersymmetric parameter space including
non-universal gaugino masses can be found in Ref.~\cite{belanger1},
and a more complete analysis seems to produce similar results to the
ones we will show in this letter~\cite{belanger2}. In general less
model-dependent analyses become increasingly promising the more hard
data becomes available~\cite{favorite}.\bigskip

\underline{In the setup of our analysis} we try to minimize the
effects of MSSM model building. Instead we start from a completely
general non-unified MSSM spectrum only assuming $R$ parity, since
without $R$ parity the LSP as a cold dark matter candidate ceases to
exist. The LSP we assume to be the lightest neutralino $\nne$. On top
of that we only assume a very minimal set of general LEP2 limits for
charginos, sleptons and sneutrinos, and the measured relic density
$0.05<\Omega_\chi h^2<0.3$. We note, however, that the assumption of a
general LEP2 mass limit implicitly assumes an $SU(2)$ relation between
the mass of the left handed slepton and the sneutrino in each
generation. We do not consider the possible effects of complex soft supersymmetry breaking terms \cite{complex}.

\section{Experimental Limits}
\label{sec:ex}

The present density of dark matter has been measured to be
$\Omega_{\rm{CDM}}h^2 = 0.12 \pm 0.04$~\cite{silk}. This result is the
combination of data from measurements of the cosmic microwave
background, type Ia supernova red-shifts, 2dFGRS and SDSS galaxy
red-shifts and data from the Hubble Space Telescope. In order to
provide a suitable dark matter candidate, a set of SUSY parameters
must yield an LSP with a relic density similar to these
observations. We will, however, consider models in which somewhat
larger or smaller densities ($0.05 < \Omega_{\rm{CDM}}h^2 < 0.30$) are
produced, acknowledging the possibility that one or more of the pieces
of the contributing cosmological evidence is not fully understood
theoretically~\cite{friedman_eq}.\smallskip

Since in this letter we want to determine an as general as possible
limit on the neutralino LSP mass, we limit ourselves to a few
experimental results which have turned out to be particularly hard to
circumvent. The single most limiting collider result is probably the
LEP2 search for supersymmetric particles such as charginos and
sleptons~\cite{lep_char,lep_slep}. It would be far beyond the scope of
this letter to discuss in detail all LEP2 limits.  Instead we require
all scalar leptons and charginos to be heavier than $103\gev$. For
particles which decay to leptons and the neutralino LSP this is a
direct experimental bound, while for sneutrinos in most cases it
requires a very basic $SU(2)$ symmetry between the supersymmetry
breaking masses of the left handed slepton and the sneutrino of one
lepton flavor.  The effect of slightly reduced mass limits (for
example from background effects at LEP which might push the mass
limits below the kinematic boundary) will be discussed later and the
numerical impact can easily be determined. Specific properties of the
MSSM spectrum are claimed to have significant effects on the mass
limits and we discuss them in greater detail in
Section~\ref{sec:cons}. Generally, the LEP2 limits become even harder
to circumvent for a very light neutralino LSP which is what we require
in the following analysis\footnote{The reach of the scalar tau search
at LEP should for example improve in a light LSP regime: for the decay
$\sle\sle^* \to \tau\bar{\tau} + \met$ two light LSPs make it easier
to distinguish the $2\nne 4\nu$ missing transverse momentum from the
$4\nu$ background. If we compare the signal to the chargino searches
$\cpe\cme \to \ell \ell' + \met$ we see that they are identical.  The
only complication would be the acceptance for low values of $\met$
which the light LSP will help to avoid.}.

Since the question how light the LSP can actually be does not directly
depend on the squark and gluino masses, we decouple these particles and
avoid their, in principle, very powerful mass
limits~\cite{tev_limits,run2report}. This choice of heavy squark
masses also means that stop--neutralino co-annihilation can be
neglected in our analysis~\cite{stop_coann}.  We also
decouple the charged Higgs boson and essentially avoid the $b \to
s\gamma$ constraints (which we still check for all our parameter
points)~\cite{bsgamma_ex,bsgamma_th}. Generally, these additional
parameters will not have a major direct impact on the minimum LSP mass
in a general non-unified MSSM. However it has been shown that the
impact through further theoretical assumptions on the MSSM spectrum
can be very significant, in particular once the mass spectrum includes
mass degeneracies.\bigskip

For light neutralinos, in particular, the limit on the invisible $Z$
decay width can become important. We require that $Z$ decays to
neutralino LSPs contribute less than one standard deviation to the
measured neutrino contribution, which agrees with the Standard Model
prediction, \ie $\Gamma_{Z\to\chi\chi}<4.2\mev$. This invisible width
limit has a major impact on another additional LEP2 search channel:
the single photon production. The Standard Model background process is
$e^+e^- \to Z^{(*)} \to \nu\nu$ with an additional photon radiated off
the initial state. The total cross section for the production of
$Z\gamma$ at LEP2 is less than $31\pb$ for a minimum $1\gev$
transverse momentum cut on the photon. The supersymmetric signal
process is the production of two lightest neutralinos and an
additional photon, which can be radiated off the incoming electrons or
the $t$ channel selectron. For light Higgsinos, the dominating process
is again $Z\gamma$ production, but with a decay of the $Z$ to
neutralinos. Hence the upper limit on the invisible $Z$ decay width
translates into a suppression factor, $S/B<8.10^{-3}$, which does not
allow for a significant signal at LEP2. Any additional initial state
radiation tends to lead to a far forward photon and we have checked
that it will not increase this approximate result by more than a
factor of two. The situation is different for light gauginos, in which
case the dominating diagram is photon radiation off the $t$ channel
selectron. We checked typical parameter points from our analysis
assuming that the selectron be heavier than $103\gev$ and we find
cross sections below $0.13\pb$. This translates into $S/B<0.02$ or
$S/\sqrt{B}<0.9$ before cuts and for irreducible backgrounds only. We
therefore conclude that the single photon channel does not pose any
obvious limits on the parameter points we find in our analysis. The
bottom line that there are no experimental limits on the neutraqlino
LSP mass without any additional constraints is in agreement with a
more complete analysis presented in the context of teh KARMEN time
anomaly~\cite{karmen}.

As described in Section~\ref{sec:dm} and as we will also see in our
analysis later, the limit on the neutralino LSP mass does only mildly
depend on the selectron mass value. In contrast a large selectron mass
alone could lead to a decoupling of the single photon signature. We
emphasize that in our analysis we do not decouple the selectrons to
respect the current experimental limits on single photon production.
All our parameter points even with a low selectron mass automatically
obey the LEP2 limits.

\section{Supersymmetric Parameter Space}
\label{sec:scan}

\begin{figure}[t] 
\begin{center}
\includegraphics[width=13.0cm]{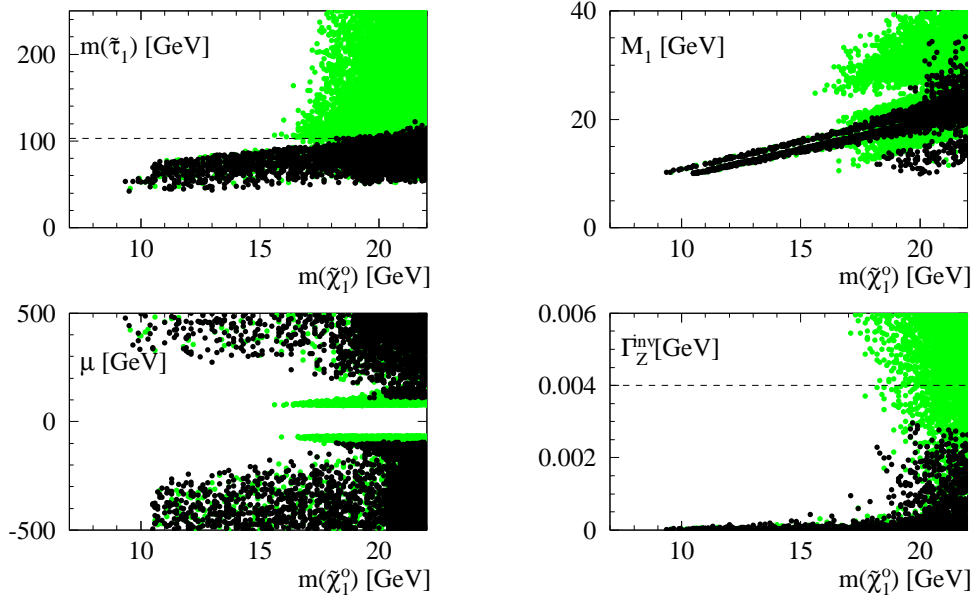}
\end{center}
\vspace*{0mm}
\caption[]{\label{fig:one} 
  The MSSM data points with the neutralino LSP mass on one axis. The
  other axis' in the four panels show the lightest slepton mass, the
  bino mass parameter $M_1$, the Higgsino mass parameter $\mu$ and the
  contribution of the decay $Z \to \nne \nne$ to the invisible $Z$
  decay width. The color coding corresponds to the light chargino
  mass with only the black points respecting 
  $\mce>103\gev$.  The dashed lines are the assumed experimental
  limits. Note that in contrast to the black points, not
  all of the green (grey) points with too small chargino masses are included
  in the frames. Moreover the black points might hide green (grey)
  points below them.}
\end{figure}

To probe the supersymmetric parameter space, we use a Monte Carlo scan
assuming that squarks, gluinos and heavy Higgs bosons are decoupled
with masses of $1\tev$. We scan over the relevant neutralino mass
parameters $M_2=50$ to $500\gev$, $|\mu|=50$ to $500\gev$ and $M_1=10$
to $40\gev$. Moreover we scan over a slepton mass parameter from
$100\gev$ to $250\gev$. As we will argue later in this section the
most relevant parameter is the lightest slepton mass, \ie the mass of
the lighter stau $\sle$.  We do scan over $\tan\beta$ but we do not
find any dependence of the neutralino LSP mass on $\tan\beta$ after
imposing all other constraints. Its impact on the light stau mass is
washed out by the simultaneous scan over $\mu$. Negative values of
$M_1$ which are often ignored and which, for example, decouple
neutralino mediated decays of sleptons or squarks~\cite{light_sbottom},
have no impact on our analysis. A second run was conducted with
similar parameter ranges, but allowing $M_1$ as large as $60 \gev$ and
the common slepton mass parameter as large as $400 \gev$. This second
set is used in Fig.~\ref{fig:three}. Last but not least, for all plots
we add $\sim 100$ data points with slepton and chargino masses right
at the LEP2 limits and very low neutralino LSP masses to model the
envelope around the lowest allowed neutralino LSP masses. We emphasize
that in all plotted parameter points, the strongly interacting MSSM
partners are assumed to be heavy and all non-tau sleptons respect the
LEP2 mass limit.\bigskip

All points in the supersymmetric parameter space allowed by the relic
density $\Omega_\chi h^2=0.05$ to $0.3$ are given in
Fig.~\ref{fig:one}.  The green (grey) points have a too small light
chargino mass $\mce<103\gev$ while the black points obey the LEP2
limit $\mce>103\gev$.  In the upper left panel of Fig.~\ref{fig:one},
we see the correlation between the light chargino and the lightest
slepton mass: once the neutralino LSP becomes very light the relic
density increases rapidly beyond the allowed limit $\Omega_\chi
h^2<0.3$. The only way to reduce this relic density is the
annihilation of two LSPs to leptons. The dominant diagram for the
annihilation of gaugino LSPs is the $t$ channel exchange of the
lightest slepton (the lighter stau) in the process $\nne\nne \to
\tau\tau$. A too large $\sle$ mass will immediately lead to an
over-closing of the universe. The effect of the stau mass limit is
shown in Fig.~\ref{fig:one}. In our analysis, we generally assume that
the lighter stau be the lightest scalar lepton. All arguments, however,
translate trivially to any other lightest slepton case.  The black
points which respect the chargino mass limit clearly prefer a light
stau, which is experimentally ruled out. Balancing the limit on the
relic density with the stau mass limit gives a minimum LSP mass of
$\mne \lesssim 18\gev$. For a fixed chargino mass limit the effect of
a relaxed stau mass limit can also be read off this figure: for
example a reduced mass limit $\mle>80\gev$ already allows an LSP mass
of $\sim 10\gev$.\bigskip

The only way to avoid the correlation described above would be a
Higgsino LSP which can annihilate through an $s$ channel $Z$ boson.
The next two panels however show how it is the bino mass parameter
$M_1$ which drives the LSP mass to low values. In the third panel of
Fig.~\ref{fig:one}, we do see two tails of light LSP masses at small
$|\mu|$ values, which we deliberately limit to $|\mu|>50\gev$. The
reason is that these point represent light Higgsino dark matter and
are firmly ruled out by the chargino mass limit. The lower right panel
of Fig.~\ref{fig:one} shows how the chargino mass works together with
the invisible $Z$ width measurement: all black parameter points
with sufficiently heavy charginos render a gaugino LSP which does not
couple to the $Z$ boson and, therefore, automatically avoids the
invisible $Z$ width bound. Once we move towards a lighter Higgsino LSP,
the invisible $Z$ decay limit is immediately violated. This shows how
the LEP2 chargino mass limit, as well as the invisible $Z$ width
measurement, firmly rule out light dark matter with a non-negligible
Higgsino content~\cite{higgsino_dm}.\bigskip

\begin{figure}[t] 
\begin{center}
\includegraphics[width=13.0cm]{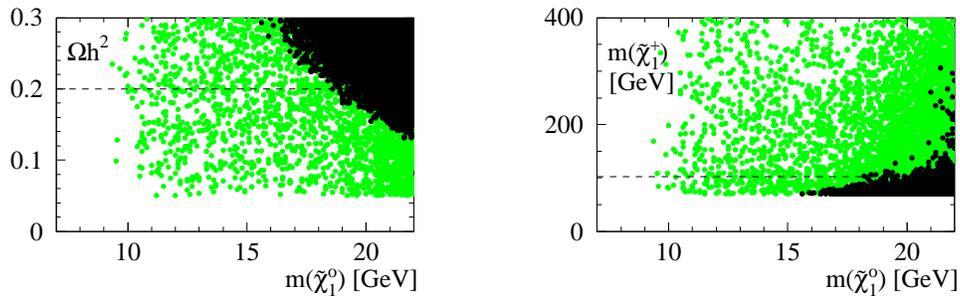}
\end{center}
\vspace*{0mm}
\caption[]{\label{fig:two} 
  The MSSM data points with the neutralino LSP mass on one axis. The
  other axis' in the two panels show the LSP relic density and the
  lighter chargino mass. The color coding corresponds to the lighter
  stau mass with only the black points respecting
  $\mle>103\gev$.  The dashed lines are the experimental limits on the
  chargino mass and a possible improved limit on the relic density.
  Note that in contrast to the black parameter points, not all of
  the green (grey) points with too small chargino masses are included in the
  frames. Moreover the black points might hide green (grey) points below
  them.}
\end{figure}

To illustrate the interplay between the chargino and the stau mass
limits, we once more print all the points in Fig.~\ref{fig:one}, but
with a different color coding: in Fig.~\ref{fig:two} the black points
obey the mass limit $\mle>103\gev$, all other points are printed in
green (grey). As described above, a heavier stau leads to an
over-closed universe, unless the LSP is a Higgsino. The behavior which
can be seen in the left panel of Fig.~\ref{fig:two} reflects the lower
limit $|\mu|>50\gev$ in our scan. The allowed black data points show a
strong correlation of the gaugino LSP mass and the relic density, this
time for a fixed limit of the light stau mass: the allowed relic
density clearly determines the minimum LSP mass under the condition
that the stau mass limit is not violated. Again, the way to obtain a
light LSP is the admixture of Higgsino content, to couple to the $s$
channel $Z$ annihilation diagram. But then small LSP masses yield a
small chargino mass, and in the right panel of Fig.~\ref{fig:two} we
again see the few points which respect both LEP2 mass limits and again
find a minimum LSP mass $\mne\lesssim18\gev$. We also see how small
LSP masses can be realized with very heavy chargino masses
$\mce\gtrsim300\gev$. This seems to be a slight asymmetry between the
chargino mass and the scalar tau mass dependence: since the $\sle$
mass directly enters the dominating neutralino LSP annihilation
diagram, the relation of its mass with the neutralino LSP mass is very
smooth, as can be seen in the first panel of Fig.~\ref{fig:one}.  The
chargino mass in contrast has to be heavier than a certain value for
any given LSP mass but its actual allowed values are not strongly
correlated with the neutralino LSP mass.\bigskip

The only non-trivial assumption in the analysis presented above is the
$SU(2)$ relation between the left handed slepton masses and the
corresponding sneutrino mass. As always, we implicitly assume that the
lighter stau be the lightest slepton, but it is obvious from the
discussion above that for very small neutralino LSP masses, all slepton
masses have to be right at the LEP limit of $103\gev$. Explicitly we
first check what happens if only one lepton generation is available in
the annihilation process, \ie if, for example, the selectrons and smuons
are completely decoupled: this decoupling changes the mass limit on
the neutralino LSP from $18\gev$ to $25\gev$, independent of which
lepton generation is available. In contrast, the decoupling of all
sneutrinos has no significant effect on the LSP mass limit. This can
be understood by comparing the different couplings $\ell \tilde{\ell}
\tilde{B}$ for left and right handed sleptons and for sneutrinos. The
sneutrino coupling is indeed suppressed. As expected from this
argument, limiting the neutralino LSP annihilation to sneutrino
mediated processes only yields an increase in the mass limit from
$18\gev$ to $27\gev$. In this sense, a slight violation of the $SU(2)$
symmetry between the masses in one lepton generation, which can yield
slightly lighter sneutrinos than the LEP2 mass bound of $103\gev$,
would not significantly change the mass limit we obtain.

\section{Conspiracies?}
\label{sec:cons}

Going beyond the generic features described in Section~\ref{sec:scan},
there might be a way of avoiding the LEP limits on charginos and
sleptons: if the LSP is only very few$\gev$ lighter than the particle
produced then the additional final state leptons become soft and the
LSP becomes slow. We briefly comment on the effect of two possible
mass degeneracies on our lower limit for the neutralino LSP
mass:\bigskip

First, the lighter chargino mass can be almost mass degenerate with the
lightest neutralino. This leads to additional neutralino--chargino
co-annihilation as a way to circumvent the impact of the slepton mass
limits. One way to achieve this degeneracy is to have $|\mu|$ define
both the lightest neutralino mass and the light chargino
mass. However, in Section~\ref{sec:scan} we learned that we violate
the invisible $Z$ decay width measurement in the case of a light
Higgsino LSP. Another way this same mass degeneracy occurs is for $M_2
\ll M_1,|\mu|$, \ie for a dominantly wino light chargino and lightest
neutralino.  Moreover, it can once more arise from a diagonal
parameter choice $M_2=M_1 \ll |\mu|$. The effects on the spectrum are
identical: the light wino-type chargino decays into slowly moving
leptons or quarks and the neutralino LSP.

If a $Z$ boson decays into two charginos which are mass degenerate
with the LSP, all decay products escape the detector unobserved and the
process contributes to the invisible $Z$ decay width. In the limit
$\mce \ll m_Z$ and for a pure wino-type chargino we can link the
partial decay width of the $Z$ boson to the decay width to one
generation of neutrinos $\Gamma(\cpe \cme) \sim 4.7 \,
\Gamma_{\nu\nu}$. The crucial observation is that while the $Z$ boson
does not couple to the gaugino fraction in a pair of neutralinos, it
does couple to the gaugino fraction in a chargino pair roughly with
the same strength as it couples to the Higgsino fraction. The
experimental limit on the invisible decay of a $Z$ boson translates
into $\Gamma_{\rm inv} \lesssim \Gamma_{\nu\nu}/40$.  The typical
finite mass correction for a $18\gev$ decay product is $\sqrt{1-4
m^2/m_Z^2} \sim 0.9$ and will not yield the required suppression by a
factor of $1/200$. We can, therefore, safely assume that mass degenerate
neutralinos and charginos can indeed escape detection for continuum
production but not for $Z$ decays.  Their masses have to be above half
of the $Z$ boson mass and are not in the very light LSP regime which
we are exploring.\medskip

The second type of mass degeneracy occurs between the neutralino LSP
and the lighter stau. This allows a very efficient annihilation of
LSPs and prevents the universe from over-closing, even for a very
light gaugino LSP. Moreover, it allows neutralino--stau
co-annihilation to reduce the relic density further~\cite{stau_coann}.
As for the charginos it is not trivial to avoid the $Z$ decay data,
since the normalization of events is known, \ie there is a limit on
invisible decays. The typical slepton partial width in the light
slepton approximation is $2 (T_3-Q s_w^2)^2 \, \Gamma_{\nu\nu}$ for a
left handed and $2 (-Q s_w^2)^2 \, \Gamma_{\nu\nu}$ for a right handed
slepton. This is, again, too large to be hidden in the error on the
invisible $Z$ decay width. However, the stau is the lightest slepton
just because it mixes the weak eigenstates into mass eigenstates and
yields a light mass eigenstate $\sle$.  This mixing can be used to
decouple the $\sle$ from the $Z$ boson, the same way that a light
sbottom can avoid the $Z$ decay limits~\cite{light_sbottom}. The tree
level coupling to the $Z$ boson vanishes for a choice of the scalar
mixing angle $\cos^2\theta = Q/T_3 s_w^2$ which, in case of the stau, is
$\theta_\tau \sim \pi/4$. This decoupling condition does not affect
the LSP annihilation cross section and, therefore, it would be possible
to have a very light neutralino and tau slepton and get the correct
amount of gaugino dark matter. The only worry is how to get one very
light tau slepton with a mass of less than $18\gev$ and keep the
second stau heavy. Mixed pairs $\sle\slz$ can in that case be produced
at LEP2 even if the heavier stau has a mass of up to $\sim
180\gev$. This mixed production cross section is proportional so $\sin
2 \theta_\tau$ and will not be suppressed around the decoupling point
for the light stau $\theta_\tau \sim \pi/4$.  At this point it is
obvious that this kind of scenario is ruled out assuming any scalar
mass unification, involving different flavor slepton masses, Higgs
masses and squark masses. Moreover, it will have to be carefully
checked that the large stau mass splitting does not violate the
experimental limits on the rho parameter.\bigskip

\begin{figure}[t] 
\begin{center}
\includegraphics[width=13.0cm]{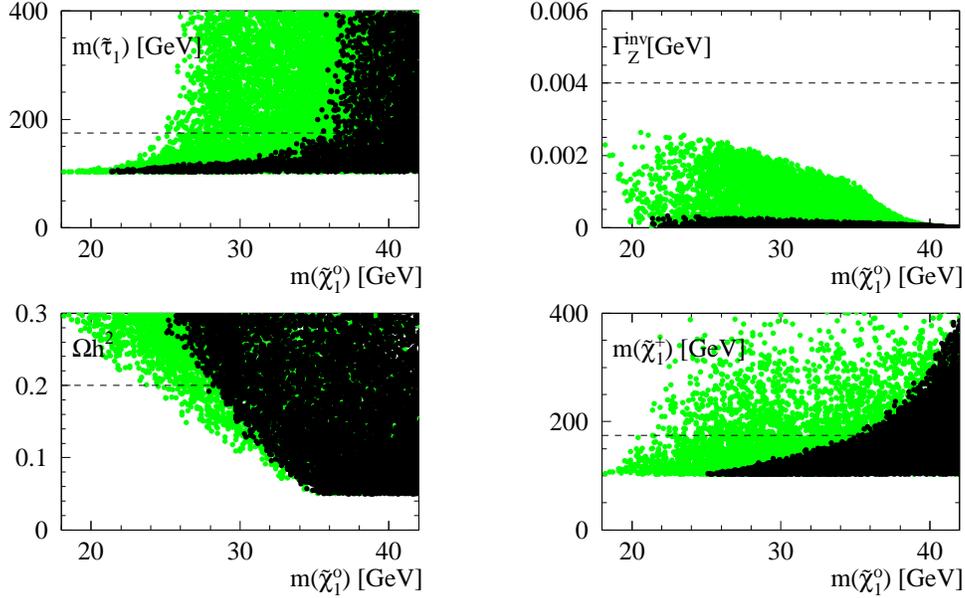}
\end{center}
\vspace*{0mm}
\caption[]{\label{fig:three} The MSSM data points with the neutralino
  LSP mass on one axis.  All black and green (grey) parameter points
  respect the $103\gev$ LEP limits as well as all other limits we
  impose.  Upper row: versus the lightest slepton mass and versus the
  $Z$ invisible decay width, like in Fig.~\ref{fig:one}. Here the
  color coding corresponds to the lightest chargino mass with the
  black points indicating $\mce>175\gev$. Lower row: versus the LSP
  relic density and versus the light chargino mass, like in
  Fig.~\ref{fig:two}. The color now coding corresponds to the lightest
  slepton mass with the black points indicating $\mle>175\gev$.}
\end{figure}

Going back to the starting point of this section, we want to stress
that the statement that an almost mass degenerate stau-neutralino
combination can escape the LEP2 trigger has to be carefully
examined. Indeed the decay products, \ie the leptons and the
neutralino LSP, will not gain any momentum from the stau or chargino
decay. However, the decaying particles, \ie the stau or the chargino,
are very light compared to the beam energy. They themselves will move
through the central detector rapidly and in turn boost their decay
products. While we are not aware of a detailed study of this part of
supersymmetric parameter space and while we did, therefore, point out
possible complications in this section we very much doubt that mass
degeneracies could hide a light stau or chargino from the LEP2
experiments. To finally close this (non-existing) loop hole should be
a simple exercise for these experiments.

\section{Outlook} 
\label{sec:concl}

Following the detailed discussion above, we emphasize that the
neutralino LSP mass limit $\mne \gtrsim 18\gev$ is only possible
because we have mass limits on charginos and all sleptons
simultaneously. One of the two alone will not constrain the general
MSSM parameter space. The kind of collider which seems to be designed
to fulfill this task of multiple searches for new particles are
$e^+e^-$ colliders. The lead in this field has by now been changed
from LEP2 to a linear collider~\cite{tesla_tdr}. The latter in a first
stage could for example collect data at the top threshold. Assuming
that this initial stage might not be sufficient to exploit the
$\nne\nnz$ production channel and discover the lightest neutralino we
estimate what a $175\gev$ limit on charginos and sleptons would mean
for the neutralino LSP mass.  The results are depicted in
Fig.~\ref{fig:three}. The upper row of plots is color coded the same
way as Fig.~\ref{fig:one}: only the black points respect the mass
limit for the lightest chargino $\mce>175\gev$. As expected the
minimal possible neutralino LSP mass decreases once we enforce the
stau mass limit $\mle>175\gev$, yielding a lower limit of
$\mne>35\gev$. For this figure, we implicitly assume that the light
stau be the lightest slepton. A minimum mass of $175\gev$, therefore,
means that all other sleptons respect this mass limit as well. The
invisible $Z$ decay width does not have any impact on this result. The
lower row of plots in Fig.~\ref{fig:three} is color coded just like
Fig.~\ref{fig:two}: only the black points respect the projected TESLA
limit on the lightest slepton $\mle>175\gev$.\smallskip

In the left panel of the lower row in Fig.~\ref{fig:three} we also see
the change of the allowed neutralino LSP mass \eg if we require the
relic density to be the central measured value $\Omega_{\rm{CDM}}h^2 =
0.12 \pm 0.04$~\cite{silk}: no LSP masses below $\sim 30\gev$ are
consistent with this central value and the $103\gev$ LEP2 bounds. An
upper limit of $\Omega_\chi h^2<0.2$ will automatically increase the
lower LSP mass limit to $\mne>25\gev$.\bigskip

\underline{In Summary} we investigate how the LEP2 limits narrow the
allowed range of the mass of a neutralino LSP in a general $R$ parity
conserving MSSM. These LEP2 mass limits have to be respected by all
scalar leptons including all sneutrinos and by the charginos. The only
assumption we use is that the LSP be responsible for the observed dark
matter density. Under these assumptions the absolute lower limit on
the neutralino LSP mass is $\mne>18\gev$. The lowest values for the
LSP mass require all sleptons and the chargino to be just above the
LEP2 limit of $103\gev$ and yield an allowed relic density at the
upper boundary of $\Omega_\chi h^2 \sim 0.3$.\bigskip

Note added: after this paper had appeared as a preprint a similar
analysis was published, which pointed out that small neutralino LSP
masses are allowed for strongly mixed gaugino--Higgsino
neutralinos~\cite{new}. The annihilation of these LSPs has to mainly
proceed through a light pseudoscalar Higgs boson $A$ in the $s$
channel. To not over-close the universe the pseudoscalar Higgs boson
mass has to be light, sitting in an allowed corner of the MSSM
parameter space with $m_A \sim 90\gev$ and $m_h \sim 90\gev$ for the
light scalar Higgs boson mass~\cite{lep-ewwg}. The main constraint on
this kind of models then becomes the invisible $Z$ decay width and
even more importantly the $b \to s\gamma$. We point out that after
including all constraints we do find smaller LSP masses when fixing
$m_A=90\gev$ and scanning over the MSSM parameter space. For a fixed
value $m_A=110\gev$ there still remain a few parameter points with an
allowed LSP mass between 16 and $17\gev$, while for $m_A \gtrsim
130\gev$ the LSP annihilation through the $s$ channel pseudoscalar is
not efficient enough to impact the results described in this letter.


\acknowledgements We would like to thank Joakim Edsjo for help on many
crucial issues. We would also like to thank Tao Han, Ulrich Nierste, 
Rohini Godbole, and Abdel Djouadi for very helpful discussions. In
particular we would like to thank Toby Falk for numerous discussions
throughout this project and for his careful reading of the
manuscript. Without him we would certainly not have been able to write
this paper. This research was supported in part by the University of
Wisconsin Research Committee with funds granted by the Wisconsin
Alumni Research Foundation and in part by the U.~S.~Department of
Energy under Contract No.~DE-FG02-95ER40896.


\bibliographystyle{plain}

\end{document}